\documentclass[prd,twocolumn,nofootinbib,showpacs]{revtex4}
\usepackage{graphicx, epsfig}
\usepackage{color}
\usepackage{mathrsfs}
\usepackage{amsmath}
\usepackage{amssymb}
\begin{document}
\newcommand{\be}{\begin{equation}}
\newcommand{\ee}{\end{equation}}
\title{Notes On The Post-Newtonian Limit Of Massive Brans-Dicke Theory}
\author{Mahmood Roshan$^1$}
\email[]{rowshan@ut.ac.ir}
\author{Fatimah Shojai$^1$}
\email[]{fshojai@ut.ac.ir}
\affiliation{$^1$ Department of Physics, University of Tehran, Tehran,
Iran}

\begin{abstract}
We consider the Post-Newtonian limit of massive Brans-Dicke theory and we make some notes about the Post-Newtonian limit of the case $\omega=0$. This case is dynamically equivalent to the metric $f(R)$ theory. It is known that this theory can be compatible with the solar system tests if Chameleon mechanism occurs. Also, it is known that this mechanism is because of the non-linearity in the field equations produced by the largeness of the local curvature relative to the background curvature. Thus, the linearization of the field equations breaks down. On the other hand we know that Chameleon mechanism exists when a coupling between the matter and the scalar field exists. In the Jordan frame of Brans-Dicke theory, we have not such a coupling. But in the Einstein frame this theory behaves like a Chameleon scalar field.
By confining ourselves to the case $\omega=0$, we show that "Chameleon-like" behavior can exist also in the Jordan frame but it has an important difference compared with the Chameleon mechanism. Also we show that the conditions which lead to the existence of "Chameleon-like" mechanism are consistent with the conditions in the Post-Newtonian limit which correspond to a heavy scalar filed at the cosmological scale and a small effective cosmological constant. Thus, one can linearize field equations to the Post-Newtonian order and this linearization has not any contradiction with the existence of "Chameleon-like" behavior.
\end{abstract}
\maketitle
\section{introduction}
The dynamics of massive Brans-Dicke (BD) theory can be determined by the following action
\begin{equation}
S=\frac{1}{2k^{2}}\int d^{4}x \sqrt{-g}[\phi R-\frac{\omega}{\phi}\partial_{\mu}\phi\partial^{\mu}\phi-V\left(\phi\right)]
+ S_{m}[g,\psi]
\label{action}
\end{equation}
where $k^{2}=8\pi G$ and $S_{m}$ is the matter action. In the original BD theory, the mass term $V\left(\phi\right)$ is zero and so the scalar field is massless. The Post-Newtonian (PN) limit of massless BD theory has been investigated in \cite{nutku}. It is known that the massless BD theory can not always produce the corresponding GR case in the limit $\omega\rightarrow\infty$. And there exist some exact solutions that doesn't go over to the Einstein's theory in this limit \cite{faraoni book}. Banerjee \textit{et al} \cite{infinit omega} argued that this is due to the zero trace of the energy-momentum tensor of matter. But even for a non zero trace of energy-momentum tensor, the limit of $\omega\rightarrow\infty$ doesn't lead to GR necessarily \cite{bahadra}. Whether the BD theory behaves like GR at $\omega\rightarrow\infty$ or not, depends on validity of the main assumption of PN formalism. For example, in the PN limit we assume that the perturbation of the scalar field due to the local gravitating system under consideration is very small compared to its cosmological background value. If this assumption fails then this theory will not behave as GR in the PN limit. The situation is more complex in the massive BD theory even if the coupling constant $\omega$ is zero. The weak field limit of this case is the subject of controversy. More specifically, several authors claimed that the metric $f(R)$ theory (which is equivalent to the massive BD theory with $\omega=0$) is characterized by an ill-defined Newtonian regime \cite{capo}.

It is known that, the Jordan frame of BD theory is related to the Einstein frame via a conformal transformation. In the Einstein frame there exists a conformal coupling between the matter and the scalar field which leads to an interaction between them \cite{faraoni book}. Hence, this theory in the Einstein frame can be considered as an interacting quintessence model\cite{CQ} . On the other hand, this interaction leads to a density-dependent scalar field mass i.e. the associated mass of the scalar field can change with environment \cite{khoury}. This behavior is the reason for naming such a theory the
Chameleon theory \cite{khoury}. The mass of the scalar field can be very large at the dense places and so the scalar field interaction with the matter can be strongly suppressed. Also, under the condition named the thin-shell condition \cite{khoury} the scalar field outside the source is produced only with a very thin shell of matter near the surface of the source. In the other words, the interior part of the source has no contribution to the generating of the scalar field.

 An important question can be arisen here. We know that some ambiguity exist for the physical equivalence of the Einstein frame and the Jordan frame. Can we conclude the existence of the Chameleon mechanism in the Jordan frame from its existence in the Einstein frame?
 We expect that one should see this behavior directly in the Jordan frame without switching to the Einstein frame. Also it is natural to expect the trace of this behavior in the PN limit. However, it is claimed in the literature that when Chameleon mechanism occurs, linearization of the field equations (which is necessary in the PN limit) breaks down and the behavior of the scalar field is governed by non-linear dynamics\cite{defelice}. But, by working directly in the Jordan frame we show that there is no Chameleon mechanism in the form of \cite{khoury} but a different version of it which we named it "Chameleon-like". We verify that linearization of the field equations to the PN order has not any contradiction with the existence of the "Chameleon-like" mechanism and the required conditions for the existence of this mechanism are consistent with the required conditions in the PN limit for the viability in the solar system.

The outline of this paper is as follows: In Sec.II we review the PN limit (expansion) of the massive BD theory and we look for non-linear terms responsible for the Chameleon mechanism. Although, we will show that such terms do not exist. In Sec.III we confine ourselves to $\omega=0$ case and show that "Chameleon-like" behavior can exist in this theory. Finally, we compare the result of PN limit and Chameleon-like behavior and we show that they are consistent with each other. Also, we clarify the range of validity of the PN expansion and show that this expansion dose not break down when the Chameleon mechanism occurs.
\section{Post-newtonian limit of massive BD theory}
In order to find some solutions of the field equations in the PN approximation, we expand the equations of motion around the background values of the metric and the scalar field. More specifically, we will use: $g_{\mu\nu}=\eta_{\mu\nu}+h_{\mu\nu}$, $g^{\mu\nu}=\eta^{\mu\nu}-h^{\mu\nu}$, $\phi\left(x,t\right)=\phi_{0}\left(t\right)+\varphi\left(x,t\right)$ and $V\left(\phi\right)=V_{0}+\varphi V_{0}'+ \varphi^{2} V_{0}''/2+...$. It is worth to mention that we are working in the Post-Newtonian coordinate system \cite{will book}. The PN limit of any scalar-tensor theory requires a knowledge of: $g_{00}$ to $O(4)$, $g_{0j}$ to $O(3)$, $g_{ij}$ to $O(2)$ and $\varphi$ to $O(4)$. Note that in the PN limit we take $v^{2}\sim U \sim O(2)$ where $v$ is the characteristic velocity of particles and $U$ is the Newtonian gravitational potential. Thus, we look for the solutions of the field equations in the form of a Taylor expansion as
\begin{eqnarray}
\begin{split}
&g_{00}\simeq-1+h^{(2)}_{00}+h^{(4)}_{00},\\
&g_{0j}\simeq h^{(3)}_{0j},\\
&g_{ij}\simeq \delta_{ij}+ h^{(2)}_{ij},\\
&\phi\simeq \phi_{0}+\varphi^{(2)}+\varphi^{(4)}.
\end{split}
\end{eqnarray}
Variation of the action \eqref{action} with respect to $g_{\mu\nu}$ and $\phi$ will yield to the following field equations respectively:

\be
\begin{split}
R_{\mu\nu}=\frac{k^{2}}{\phi}(T_{\mu\nu}-&\frac{1}{2}g_{\mu\nu}T)+\frac{\omega}{\phi^{2}}\partial_{\mu}\phi\partial_{\nu}\phi
+\frac{1}{\phi}\nabla_{\mu}\partial_{\nu}\phi\\
&+\frac{1}{2\phi}g_{\mu\nu} [\Box\phi+V\left(\phi\right)]
\label{me}
\end{split}
\ee
\be
\Box\phi -\frac{dV_{eff}}{d\phi}=\frac{k^{2}}{3+2\omega}T
\label{se}
\ee
\be
\frac{dV_{eff}}{d\phi}=\frac{1}{3+2\omega}\left(\phi \frac{dV}{d\phi}-2V\right)
\label{sc}
\ee
The full derivation of the PN limit of the massive BD theory can be found in \cite{olmo}. Let us just review in brief the results of \cite{olmo} which are necessary for our work. The field equation \eqref{se} in the PN approximation is
\be
\left[\nabla^{2}-m_{0}^{2}\right]\varphi^{(2)}(x,t)=-\frac{k^{2}\rho}{3+2\omega}.
\ee
Where $m_{0}^{2}$ is the mass associated with the scalar field at the cosmological scales given by
 \be
m_{0}^{2}=\frac{\phi_{0}V_{0}''-V_{0}'}{3+2\omega}.
 \ee
 in which prime denotes differentiation with respect to $\phi$.  In the Solar system the expression of $h^{(2)}_{00}$ and $h^{(2)}_{ij}$ take the following forms far from the source
\begin{eqnarray}
\begin{split}
&h^{(2)}_{00}=2\frac{G_{eff}M_{\odot}}{r}+\frac{\Lambda_{eff}}{3}r^{2},\\
&h^{(2)}_{ij}=\delta_{ij}\left[2\gamma \frac{G_{eff}M_{\odot}}{r}-\frac{\Lambda_{eff}}{3}r^{2}\right].
 \label{h00}
\end{split}
\end{eqnarray}
where $\Lambda_{eff}=V_{0}/2\phi_{0}$ and
\be
G_{eff}=\frac{k^{2}}{8\pi\phi_{0}}\left(1+\frac{e^{-m_{0}r}}{3+2\omega}\right)
\ee
Note that in the derivation of equation \eqref{h00} we have used the assumption $m_{0}^{2}>0$. Using the equation \eqref{h00}, the PPN parameter $\gamma$ is given by
\be
\gamma=\frac{h^{(2)}_{ii}}{h^{(2)}_{00}}=\frac{3+2\omega-e^{-m_{0}r}}{3+2\omega+e^{-m_{0}r}}
\label{gamma}
\ee
See also \cite{perivolaropolous} in which a special potential has been investigated in the PN limit. If the scalar filed is very light then the parameter $\gamma$ is space-independent and as we expect, the massive BD theory behaves like massless BD theory i.e. this parameter takes the form $\gamma=(1+\omega)/(2+\omega)$.

 We are interested in the case $\omega=0$ which, as mentioned before, is dynamically equivalent to the metric $f(R)$ theory. In this case if the scalar field is very light then $\gamma=0.5$ which is in patent disagreement with observation since $\gamma_{ob}\simeq1$ \cite{bertotti}. However, if
 \be
 m_{0}^{2}L^{2}\gg1
 \label{condition1}
 \ee
 where $L$ represents a typical experimental length scale (below the planetary scale), then the scalar field is heavy and its interaction is short-range and it will be hidden from the local experiments. In this case $\gamma$ is near to $1$. On the other hand, the cosmological constant term $(V_{0}/6\phi_{0})r^{2}$ must be very small. Otherwise, this term will change the gravitational dynamics of local systems such as solar system. Thus, for having an observationally acceptable theory (at least in the solar system) the following condition should be satisfied too
 \be
 \frac{V_{0}}{\phi_{0}}L_{L}^{2}\ll1
 \label{condition2}
 \ee
 where $L_{L}$ is a length scale of the same order or larger than the solar system. Thus, the massive BD theory with $\omega=0$ can be compatible with the solar system tests if both conditions \eqref{condition1} and \eqref{condition2} are satisfied. Note that this conclusion are also true for metric $f(R)$ theory \cite{olmo}.

Some notes are in order here. It is claimed in the literature that the Chameleon behavior is due to the non-linear terms in the field equations\cite{defelice}. So, we have to reconsider the PN limit with more care. May be there exists some non-linear terms in the PN expansion which have been forgotten in the above considerations. The main assumptions on the scalar field in obtaining the PN
 limit are $\left|\frac{\varphi^{(2)}}{\phi_{0}}\right|\ll 1$ and $\phi\simeq\phi_{0}+\varphi^{(2)}+\varphi^{(4)}$. Let us examine the first assumption. Can we use this assumption in the gravitational systems such as solar system? In fact, $\varphi^{(2)}$ is the contribution of the scalar field due to the local gravitating system. A comparison with the results of \cite{chiba} may be useful here. In order to find the weak field limit of the metric $f(R)$ theory, Chiba, T. \textit{et al} \cite{chiba} have expanded the Ricci scalar as $R=R_{0}+ R_{1}(r)$, where $R_{0}$ is the background curvature and $R_{1}(r)$ is the perturbation produced by a given source. In order to linearize the field equations they have used $R_{1}(r)/R_{0}\ll1$. However, as we will discuss in the next section, this assumption is experimentally unacceptable near the Earth or at the solar system and therefore leads to a wrong weak field limit. Similarly, in the PN limit we linearize the field equation of $\phi$ using the condition $\left|\frac{\varphi^{(2)}}{\phi_{0}}\right|\ll 1$. If, for an arbitrary system, $\varphi^{(2)}$ is the same order or greater than $\phi_{0}$ then it is straightforward to show that the PN expansion fails and we will not able to use this approximation for that system. In this paper we do not want to find under which conditions this assumption is applicable, but we will show that when the thin-shell condition satisfied then $\left|\frac{\varphi^{(2)}}{\phi_{0}}\right|\ll 1$ is also satisfied. Thus we expect that the physical results of Chameleon-like behavior and the PN limit be similar.

 Now, consider the second assumption. We know that the odd-order terms $O(1), O(3)$ can not exist in the $g_{00}$ and $g_{ij}$ component of metric. In fact, conservation of rest mass prevents terms of order $O(1)$ and conservation of energy in the Newtonian limit prevents terms of order $O(3)$. Appearance of other odd-order terms is theory dependent, for example, $O(5)$ can not exist in GR (because of conservation of energy in the PN limit) but $O(7)$ can exist \cite{will book}. Also, we find the appropriate expansion of the scalar field from its field equation. For example, consider the field equation of $\phi$ field in the original BD theory
 \be
 \Box\phi=\frac{8\pi}{3+2\omega}T
 \ee
 By expanding the RHS for a perfect fluid to the fourth order and using an appropriate gauge condition, we can easily show that only even-order terms can appear in the expansion of $\phi$ (i.e. $\phi\simeq\phi_{0}+\varphi^{(2)}+\varphi^{(4)}$) \cite{weinberg}. However, in the massive BD theory the situation is not so trivial. In this case, consider equation \eqref{se} and assume the following expansion for the scalar field $\phi$
 \be
 \phi\simeq \phi_{0}+\varphi^{(1)}+\varphi^{(2)}+\varphi^{(3)}+\varphi^{(4)}+...
 \ee
It is an easy job to verify that if $n$ is even(odd) then $\Box\varphi^{(n)}$ contains only even(odd)-order terms. Also, we know
\be
g^{\mu\nu}\Gamma_{\mu\nu}^{\gamma}\simeq\eta^{\gamma\sigma}\left(h^{\nu}_{\sigma,\nu}-\frac{1}{2}h^{\nu}_{\nu,\sigma}\right)+ O(5)
\ee
and will use the following gauge conditions
\be
\begin{split}
&h^{\nu}_{0,\nu}-\frac{1}{2}h^{\nu}_{\nu,0}=\mathfrak{R}_{0},\\
&h^{\nu}_{j,\nu}-\frac{1}{2}h^{\nu}_{\nu,j}=\mathfrak{R}_{j}.
\end{split}
\ee
Note that for our purpose it is not necessary to fix the explicit form of $\mathfrak{R}_{0}$ and $\mathfrak{R}_{j}$. However, we should keep in our mind that $\mathfrak{R}_{0}$ is of order $O(3)$ and $\mathfrak{R}_{j}$ is of order $O(2)$. Using these gauge conditions and arranging the terms with the same orders in \eqref{se}, we obtain
\be
\begin{split}
&(\nabla^{2}-m_{0}^{2})\varphi^{(1)}=0,\\
&(\nabla^{2}-m_{0}^{2})\varphi^{(2)}=-\frac{k^{2}\rho}{3+2\omega}+\frac{\phi_{0}V_{0}'''}{2(3+2\omega)}[\varphi^{(1)}]^{2},\\
&(\nabla^{2}-m_{0}^{2})\varphi^{(3)}=(\partial_{0}\partial_{0}+h^{ij}\partial_{j}\partial_{i}+ \mathfrak{R}_{j}\partial_{j})\varphi^{(1)}+\\&\frac{1}{3+2\omega}\left[\phi_{0}V_{0}'''\varphi^{(1)}\varphi^{(2)}+\frac{\phi_{0}V_{0}''''+V_{0}'''}{6}
[\varphi^{(1)}]^{3}\right],\\
&\nabla^{2}\varphi^{(4)}=\frac{8\pi GT^{(4)}}{3+2\omega}+(\partial_{0}\partial_{0}+h^{ij}\partial_{i}\partial_{j}+\mathfrak{R}_{j}\partial_{j})\varphi^{(2)}.
\end{split}
\ee

Thus, it is obvious that the equation of motion of $\phi$, unlike the massless BD theory, allows the odd-order terms. Also, it is interesting that there exists a nonlinear term in the equation of $\varphi^{(2)}$. If $\varphi^{(1)}\neq0$ then this term will change the solutions of the metric components and also the scalar filed itself. But, we should take into account the metric field equations too. If one writes the 0-0 component of \eqref{me} to the second order, then there will be a term with the first order which should be zero separately
\be
(\nabla^{2}+V_{0}')\varphi^{(1)}=0
\ee
Similarly, from i-j component of \eqref{me} we have $\partial_{i}\partial_{j}\varphi^{(1)}=0$ and from its 0-j component $\partial_{0}\partial_{j}\varphi^{(1)}=0$. Thus, the metric field equations force $\varphi^{(1)}$ to be zero. This conclusion is also true for $\varphi^{(3)}$ and one can check it by writing equation \eqref{me} to the higher orders. As a final result of this section, one can be sure that the PN  expansion obtained in \cite{olmo} is complete and there is not any non-linear term in the field equations. So, we should see the trace of Chameleon mechanism in this PN limit. In the next section we show that "Chameleon-like" behavior can exist in this theory and then we show that how this mechanism and PN limit are connected.

\section{Chameleon-like behavior in the massive BD theory with $\omega=0$}
In this section we show that the Chameleon-like behavior can exist in the massive BD theory with vanishing coupling constant. Our main purpose is to address the question that: Can we see the Chameleon mechanism in the PN limit?

 As we discussed in the introduction, for Chameleon scalar fields the associated mass is dependent to the matter density of the environment and so the scalar field is short-range in the dense places and is long-range in the low densities such as cosmos \cite{khoury}. The main reason for such behavior is that in these theories the scalar field interacts directly with matter particles through a conformal coupling of the form $e^{\beta\phi/M_{pl}}$. On the other hand, BD theory in the Einstein frame behaves like an interacting quintessence with $\beta=-1/\sqrt{6+4\omega}$ \cite{faraoni book}. More specifically, when the coupling constant is zero, the conformal coupling takes the form $e^{-\phi/\sqrt{6}M_{pl}}$ i.e. $\beta=-1/\sqrt{6}$. Thus, we expect that in the Einstein frame, the massive BD theory behaves as a Chameleon theory, however, is it true in the Jordan frame? It is worthy to stress that the physical results can be completely different in these frames and we should be very careful in interpreting the meaning of them. Although, it is not yet clear that which one of these frames are really physical but it is usually convenient to work in the Einstein frame which is mathematically simpler and then convey the results to the Jordan frame. For more details see \cite{faraoni book} chapter 2.

As we mentioned before, the metric $f(R)$ theory is dynamically equivalent to the massive BD theory. In the context of this theory, several authors claimed that an ill-defined behavior in the Newtonian limit exists ($\gamma=0.5$)\cite{capo}. This conclusion can be inferred directly from analogy with BD theory ($\omega=0$). However, using the Chameleon mechanism in the Einstein frame, Faulkner, T. \textit{et al} \cite{faulkner} have shown that the metric $f(R)$ gravity can lead to $\gamma\sim1$. They have transformed their results to the Jordan frame.

Here, using exactly the same method of \cite{hu}, we want to explore the Chameleon behavior and the thin-shell effect in the Jordan frame of the massive BD theory with $\omega=0$. Assuming that the pressure is zero ($p=0$), one can rewrite equation \eqref{se} as
\be
3\Box\phi+2V-(\phi-1)\frac{dV}{d\phi}=R-k^{2}\rho
\ee
Note that $R=\frac{dV}{d\phi}$. Now, if $\phi\sim1$ and $\left|\frac{V}{V'}\right|\ll 1$ then the above equation can be written as
\be
3\Box\phi\simeq R-k^{2}\rho
\ee
Thus, this theory will behaves like GR if $3\left|\Box\phi\right|\ll k^{2}\rho$. In the weak field limit, where one can use $\Box\sim\nabla^{2}$, This condition can be expressed as follows
\be
\lambda_{\phi}^{2}|_{R=k^{2}\rho} \nabla^{2}\rho\ll \rho
\label{compton}
\ee
where $\lambda_{\phi}=1/m_{0}$ is the Compton wavelength of the field. This condition is the Compton condition. In order to interpret this condition, consider a source which its density changes slowly with radius say $\rho\sim r^{\varepsilon}$ (this is the case for the Earth and the Sun). Then the Compton condition can be written as
\be
\lambda_{\phi}|_{R=k^{2}\rho} \frac{\partial\rho}{\partial r}\ll \rho
\ee

Thus, if $\Delta r$ is a length scale in which the density changes concretely ($\Delta \rho\sim\rho$) then we can infer that the density changes on the length scales that are much longer than the Compton wavelength. Physically it means that the scalar field interaction will suppress on scales larger than $\lambda_{\phi}$.

As in \cite{hu}, in order to find the corresponding thin-shell condition \cite{khoury}, we take the static spherically symmetric metric around a given source at the origin as follows
\be
ds^{2}=-[1-2A(r)+2B(r)]dt^{2}+[1+2A(r)](dr^{2}+r^{2}d\Omega)
\ee
We will assume that $|A(r)|\ll1$ and $|B(r)|\ll1$ near the source and inside it. By taking into account these assumptions and the definition of the Ricci tensor, one can verify that
\be
\nabla^{2}(A+B)\simeq-\frac{1}{2} R
\label{A}
\ee
\be
\nabla^{2}B\simeq-\frac{1}{2}(R^{0}_{0}+\frac{R}{2})
\label{B}
\ee
By substituting $R^{0}_{0}$ from 0-0 component of equation \eqref{me} into \eqref{B}, we obtain
\be
\nabla^{2}B(r)\simeq -\frac{1}{3}(R-\frac{k^{2}\rho}{\phi})-\frac{V}{12\phi}
\ee
On the other hand, by using the conditions $\phi\simeq1$ and $|\frac{V}{V'}|\ll1$, we get
\be
\begin{split}
&\nabla^{2}B(r)\simeq -\frac{1}{3}(R-k^{2}\rho),\\
&\nabla^{2}A(r)\simeq -\frac{1}{6}(R-k^{2}\rho)-\frac{1}{2}k^{2}\rho.
\end{split}
\ee
And the field equation of the scalar field takes the form
\be
\nabla^{2}\phi\simeq\frac{1}{3}(R-k^{2}\rho)=-\nabla^{2}B(r)
\ee
Thus, assuming that $B(r)$ is finite at the origin, the solution of $B(r)$ and the scalar field are related as
\be
B(r)=\phi_{0}-\phi(r)=-\Delta\phi(r)
\label{bphi}
\ee
where, as before, $\phi_{0}$ is the background value of the scalar filed. Following \cite{hu}, it is convenient to define an effective mass as
\be
m_{eff}=\int\left(\rho(r')-\frac{R(r')}{k^{2}}\right)dv'
\ee
Thus, the solutions of $A$ and $B$ become
\be
\begin{split}
B(r)=-\frac{2G}{3}&\int\frac{[\rho(r')-\frac{R(r')}{k^{2}}] dv'}{|\vec{r}-\vec{r'}|}\simeq-\frac{2G}{3}\frac{m_{eff}}{r},\\
&A(r)\simeq-\frac{G m}{r}-\frac{G m_{eff}}{3 r}.
\label{B2}
\end{split}
\ee
where $m=\int\rho(r')dv'$. Finally, the PPN parameter $\gamma$ is found to be
\be
\gamma=\frac{A}{A-B}\simeq\frac{3m+\frac{1}{3}m_{eff}}{3m+m_{eff}}
\ee
It is clear form this equation that if $m_{eff}\ll m$ then $\gamma\rightarrow1$. An important result may be obtained from \eqref{B2}
\be
|B(r)|\leq\frac{2G}{3}\int\frac{\rho(r')dv'}{|\vec{r}-\vec{r'}|}=\frac{2}{3}\Phi_{N}(r)
\ee
Where $\Phi_{N}(r)$ is the Newtonian potential. Now, using equation \eqref{bphi} we get
\be
|\Delta\phi|\leq\frac{2}{3}\Phi_{N}(r)
\ee

This condition sets an upper bound on the difference between the values of the scaler field from the interior to the exterior of the source. Note that if $|\Delta\phi|\ll\frac{2}{3}\Phi_{N}(r)$ then $m_{eff}\ll m$, so $R\rightarrow k^{2}\rho$  and we will recover the GR results. This condition is called this the thin-shell condition. It is easy to convert this condition to the corresponding Einstein frame version. Then it will be exactly the thin-shell condition obtained in \cite{khoury} with $\beta=-1/\sqrt{6}$. As has been discussed in \cite{khoury,hu}, if this condition satisfied then the exterior field is only generated by the thin-shell near the surface of the spherical source. However, it is worthy to note that we have named this behavior "Chameleon-like" here because it is different from what is known in the original Chameleon theories \cite{khoury}. In fact, in the Jordan frame there is no direct interaction between the matter and the scalar field and it looks like that we have strongly restricted the scalar field evolution such that it behaves like GR. Also, we will show that if the thin-shell condition satisfied then the scalar field needs to be heavy at the cosmological scales. But, we know that in the original Chameleon theory \cite{khoury}, the thin-shell condition do not makes any restriction on the interaction range of the scalar field far away from the source. It is important to stress that although in the Jordan frame there is not any coupling between the matter and the scalar field but the scalar field couples directly to the Ricci curvature and consequently mass of the scalar field can depend to the curvature. Thus, we expect something similar to Chameleon effect in the Jordan frame. However, this "Chameleon-like" behavior ensures us that the massive BD theories with vanishing coupling constant (or equivalently metric $f(R)$ theory) can be compatible with the solar system tests. Also, in the cosmological considerations it can be different and distinguishable from $\Lambda$CDM. For example, phantom division can occur in it \cite{hu}.

Now, we are in a position to answer the main question of our paper. Is it allowable to linearize the scalar field equation when the Chameleon-like behavior exists? From the thin-shell condition and the PN expansion of the scalar field we have
\be
|\varphi^{(2)}|\ll\Phi_{N}(r).
 \label{1}
\ee
On the other hand, For having the Chameleon-like behavior, $\phi\sim1$, hence by using \eqref{1} we get
\be
\left|\frac{\varphi^{(2)}}{\phi_{0}}\right|\ll1
\ee

 Thus, as discussed before, the PN limit is applicable. In the following, we show that the second necessary condition for "Chameleon-like" behavior , i.e. $\vert V'/V\vert\gg1$, forces the scalar field to be heavy at the cosmological scales. To do this We write the Ricci scalar as follows
 \be
 R(r)=\frac{dV}{d\phi}
 \ee
Therefore, the background curvature scalar is $R_{0}=V_{0}'$. Regarding that $\phi_{0}$ is the minimum of the effective potential $V_{eff}$ and also defining $\eta(\phi)$ and $M^{2}$ as
 \be
 \eta(\phi)=\frac{V'(\phi)}{V(\phi)} \ \ \  , \ \ \  M^{2}=\frac{m_{0}^{2}}{H_{0}^{2}\phi_{0}^{2}}
\ee  
 we have
 \be
\begin{split}
&\eta_{0}=\eta(\phi_{0})=\frac{2}{\phi_{0}} ,\\
&\eta_{0}'=- \frac{2}{\phi_{0}^{2}}+\frac{M^{2}}{2} ,\\
&\eta_{0}''=\frac{4}{\phi_{0}^{3}}-\frac{3 M^{2}}{\phi_{0}}+\frac{2V_{0}^{(3)}}{\phi_{0}V_{0}'} ,\\
&\eta_{0}^{(3)}=-\frac{12}{\phi_{0}^{4}}+\frac{18 M^{2}}{\phi_{0}^{2}}-\frac{16V_{0}^{(3)}}{\phi_{0}^{2}V_{0}'}-\frac{3M^{4}}{4}+\frac{2V_{0}^{(4)}}{\phi_{0}V_{0}'},\\
&\eta_{0}^{(4)}=\frac{48}{\phi_{0}^{5}}-\frac{120 M^{2}}{\phi_{0}^{3}}+\frac{120V_{0}^{(3)}}{\phi_{0}^{3}V_{0}'}+\frac{15M^{4}}{\phi_{0}}-\frac{20V_{0}^{(4)}}{\phi_{0}^{2}V_{0}'} - \\
&\ \ \ \ \ \ \ \ \  \frac{10 M^{2}V_{0}^{(3)}}{\phi_{0}V_{0}'}+\frac{2V_{0}^{(5)}}{\phi_{0}V_{0}'},\\
\end{split}
\label{eti}
\ee
Expanding $\eta(\phi)$ in powers of $\Delta\phi$ and using \eqref{eti}, we get
\be
\begin{split}
\eta(\phi)=&\frac{V'}{V}\simeq\frac{2}{\phi_{0}}+\frac{M^{2}}{2}\Delta\phi+\frac{2V_{0}^{(3)}}{\phi_{0}V_{0}'}\frac{\Delta\phi^{2}}{2!}\\&(\frac{2V_{0}^{(4)}}{\phi_{0}V_{0}'}-\frac{3M^{4}}{4})\frac{\Delta\phi^{3}}{3!}\\&+(\frac{2V_{0}^{(5)}}{\phi_{0}V_{0}'}-\frac{10 M^{2}V_{0}^{(3)}}{\phi_{0}V_{0}'})\frac{\Delta\phi^{4}}{4!}+ \textit{O}(\Delta\phi^{5})
\end{split}
\label{eta}
\ee
Note that we have expanded $\eta(\phi)$ to the higher orders of $\Delta\phi$ because when the Chameleon-like behavior occurs  the difference between $\vert\eta(\phi)\vert\gg1$ and $\eta_{0}\sim2$ is very large. First assume that $M^{2}\ll1$, then one can rewrite $\eta(\phi)$ as
\be
\eta(\phi)\simeq \frac{2}{\phi_{0}}(1+\frac{V_{0}^{(3)}}{V_{0}'}\frac{\Delta\phi^{2}}{2!}+\frac{V_{0}^{(4)}}{V_{0}'}\frac{\Delta\phi^{3}}{3!}+\frac{V_{0}^{(5)}}{V_{0}'}\frac{\Delta\phi^{4}}{4!}+...)
\ee
For $\vert\eta(\phi)\vert\gg1$ it is necessary that

\be
\left\vert\frac{V_{0}^{(3)}}{V_{0}'}\right\vert\frac{\Delta\phi^{2}}{2!}+\left\vert\frac{V_{0}^{(4)}}{V_{0}'}\frac{\Delta\phi^{3}}{3!}\right\vert+\left\vert\frac{V_{0}^{(5)}}{V_{0}'}\right\vert\frac{\Delta\phi^{4}}{4!}\gg1
\ee
There are many cases for which this inequality is satisfied. For example, assume that these terms are of the same order, i.e.
\be
\left\vert\frac{V_{0}^{(4)}}{V_{0}^{(3)}}\right\vert\sim  \left\vert\frac{V_{0}^{(5)}}{V_{0}^{(4)}}\right\vert\simeq (\Delta\phi^{-1})\gg1
\label{see}
\ee

Thus it is ,in principle, possible to satisfy $\vert\eta(\phi)\vert\gg1$ while $m_{0}^{2}/H_{0}^{2}\ll1$. However, some notes are in order here.

Does $m_{0}^{2}/H_{0}^{2}\ll1$ guaranty that the scalar field is long range? Consider the field equation of the scalar field \eqref{se}. We rewrite it as follows
\be
\Box\phi -m_{0}^{2}(\phi-\phi_{0})+\frac{\phi_{0}V_{0}^{(3)}}{6} (\phi-\phi_{0})^{2}+...=-\frac{k^{2}\rho}{3}
\ee
If the second term is larger than the self-interacting terms, then the solution of $\phi$ for an isolated gravitating system contains the Yukawa-like term $e^{-m_{0}r}/r$. So, the magnitude of $m_{0}$ is directly related to the range of the scalar field in the sense that if the scalar field is light (heavy) then it is longe (short) range. On the other hand, if the second term is negligible compared with the third term, even for small $\Delta\phi$ , then the dynamics of the scalar field is dominated by the third term and there is not necessarily a Yukawa-like solution containing $m_{0}$ in it. In the other words, in this case, the lightness of the scalar field dose not necessarily mean the long rangeness of it. But for the Chameleon scalar fields \cite{khoury}, the lightness (heavyness) of the scalar field implies the long (short) rangeness of it. And of course, this is one of the main goals of that theory.

The second problem for the Chameleon-like behavior by the light scalar fields is that we encounter with non-smooth varying potentials, see \eqref{see}. Thus, it is very unlikely to produce the Chameleon-like behavior by the light scalar fields ($m_{0}^{2}/H_{0}^{2}\ll1$).

Now, assume that $M^{2}\gg1$. In this case one can rewrite \eqref{eta} as
\be
\begin{split}
\eta(\phi)\simeq &\frac{2}{\phi_{0}}+\frac{M^{2}}{2}\Delta\phi+ \frac{2}{\phi_{0}}(\frac{V_{0}^{(3)}}{V_{0}'}\frac{\Delta\phi^{2}}{2!}+\frac{V_{0}^{(4)}}{V_{0}'}\frac{\Delta\phi^{3}}{3!}\\&+\frac{V_{0}^{(5)}}{V_{0}'}\frac{\Delta\phi^{4}}{4!})-\frac{10 M^{2}V_{0}^{(3)}}{\phi_{0}V_{0}'}+\textit{O}(\Delta\phi^{5})
\end{split}
\ee

It is clear that without any restrective conditions  such as \eqref{see} on the form of the potential and only with the assumption that $m_{0}^{2}/H_{0}^{2}\gg1$  we can produce $\vert\eta(\phi)\vert\gg1$. In this case one can write the Ricci scalar as follows

 \be
 \frac{ R(r)}{R_{0}}\simeq 1+ \frac{m_{0}^{2}}{4H_{0}^{2}} \varphi^{(2)}(r)+\textit{O}(\Delta\phi^{2})
 \label{deltar}
 \ee
 Where $H_{0}$ is the Hubble parameter. Experimentally we know that $\frac{\Delta R}{R_{0}}=\frac{R-R_{0}}{R_{0}} \gg1$ in the solar system (it is obvious in GR). For example on the Earth $\frac{\Delta R}{R_{0}}\sim O(\rho_{Air}/H_{0}^{2}M_{pl}^{2})\sim10^{27}$ and in the solar system we can estimate this ratio using the local Dark matter density which yields $\frac{\Delta R}{ R_{0}}\sim O(\rho_{DM}/H_{0}^{2}M_{pl}^{2})\sim10^{6}$. Consequently, we can again infer from \eqref{deltar} that
 \be
 \frac{m_{0}^{2}}{H_{0}^{2}}\gg1
\ee
which can be consistent with the condition \eqref{condition1}. This condition shows that the Compton wavelength is very smaller than the Hubble length scale. This means that the scalar field is heavy in the cosmological scales. For example, near the Earth  $\Delta R/R_{0}\sim10^{27}$. By assuming that $L\sim R_{Earth}$ and $\varphi^{(2)}\sim 10^{-14}$ (for this assumption we have used \eqref{1}) and using equations \eqref{1} and \eqref{deltar} it is easy to show that $m_{0}^{2}L^{2}\sim 10^{3}$. It is clear from \eqref{deltar} that if the scalar field is very light in the cosmological scales then a gross violation of experiment will occur. We saw this result also in the PN limit where $\gamma$ was equal to 0.5 if $m_{0}^{2}L^{2}\ll1$. 

Thus, it seems like that the Chameleon-like behavior can olny occur for the heavy scalar fields ($m_{0}^{2}/H_{0}^{2}\gg1$). And this is completely against the Chameleom scalar fields \cite{khoury}. As we mentioned before, the Chameleon scalar fields are light and free at the cosmological scales.

Also since $\phi_{0}$ is the minimum of the effective potential, so we can write

\be
\frac{V_{0}}{\phi_{0}}=\frac{V_{0}'}{2} =\frac{ k^{2}\rho_{0}}{2}
\label{10}
\ee
 Where $\rho_{0}\sim 10^{-29} h^{2}\frac{g}{cm^{3}}$ is the current matter density of the universe. By multiplying this equation with  $L_{L}^{2}$, $L_{L}$ is a length scale of the same order of the solar system i.e $L_{L}\sim10^{15} cm$, we can rewrite \eqref{10} as 
 \be
\frac{V_{0}}{\phi_{0}}L_{L}^{2}\ll 1
\ee
 which is equivalent to the condition \eqref{condition2} which came from the PN limit considerations.

 So, we see that the trace of Chameleon-like behaviour is clear in the PN limit as we expected. As a final result, the PN limit of massive BD theory (with $\omega=0$) is applicable for the gravitational systems such as solar system where the Chameleon-like behaviour can occur. And one can linearise the field equations to the PN order and the result of this expansion and the Chameleon behaviour is similar. As we mentioned before the Chameleon-like behaviour which exists in the Jordan frame is different from the original Chameleon theory \cite{khoury}. On the other hand, we showed that the required conditions for this mechanism and those coming from PN limit considerations for having a viable theory in the solar system are consistent with each other.

\section{conclusion}
In this paper we have shown that although the field equation of the scalar field in the massive BD theory allows the existence of the odd-order terms in the PN expansion of the scalar filed but they are all zero. this means that there is no non-linear terms in the field equations. This is an interesting result because unlike the claim existed in the literature one can linearise field equation to the PN order and also see the Chameleon-like behaviour. It is worth to mention that our conclusion dose not mean that the non-linearity in the expansion of the Ricci scalar, coming from the largeness of the local Ricci scalar relative to the cosmological background Ricci scalar, is not important. This means that smallness of $\varphi^{(2)}$ relative to $\phi_{0}$  dose not necessarily imply $\Delta R/R_{0}\ll1$ and one should note that the condition $\Delta R/R_{0}\gg1$ has been saved in the obtaining of the PN approximation. For the massive BD theory for which the coupling constant is zero, we showed that the Chameleon-like behaviour can exist if $\phi\simeq1$, $|V/V'|\ll1$ and if the thin-shell condition $|\varphi^{(2)}|\ll\Phi_{N}(r)$ satisfied and more importantly this mechanism is different from the Chameleon mechanism. In fact, if the scalar field is forced to be heavy at the solar system through the Chameleon-like behavior then it has to be also heavy at the cosmological scales. And this is compeletly against the nature of the Chameleon scalar fields. Also, we showed that the results of PN limit and Chameleon-like behaviour are similar and when the Chameleon-like behaviour exists, the perturbation of $\phi$ due to the local system is very smaller than the cosmological background value. Thus, we can linearise the field equations to the PN order.

We finish the conclusion by two example which confirm the consistency of the PN consideration and the Chameleon-like behaviour.
Consider the model $f(R)\thicksim R^{1+\delta}$. This model can be consistent with the solar system observations if $\delta\sim-1.23\mp2.05\times 10^{-17}$ \cite{barrow} (Albeit we have taken into account the term $(1-1/\sqrt{1-e^{2}})^{-1}$ which is absent in the equation (83) of \cite{barrow}) thus Chameleon mechanism occurs for this model. On the other hand from, \eqref{deltar} and the amount of $\Delta R/R_{0}$ in the solar system, we can write
\be
\frac{m_{0}^{2}}{H_{0}^{2}}>10^{12}
 \label{end}
\ee
 By using this condition for the above model we reach to $\delta<10^{-12}$. $\delta\sim10^{-12}$ is sufficient for being consistent with the light deflection experiments, but for consistency with other observations such as the perihelion precession of Mercury, smaller amount of $\delta$ is needed. So, the condition \eqref{end} leads to a right bound on $\delta$ in agreement with the result of \cite{barrow}. For another example consider the model $f(R)\sim R-\lambda R_{c} (\frac{R}{R_{c}})^{p}$\cite{li}. The bound on $p$ coming from the Chameleon effect is $p<10^{-10}$\cite{capo2}. On the other hand, using equation \eqref{end} one can easily verify that $p<10^{-12}$ which is consistent with the previous bound. 

\section{acknowledgements} This work is partly
supported by a grant from university of Tehran and partly by a
grant from center of excellence of department of physics on the
structure of matter.

\end{document}